\documentclass[twocolumn,showpacs,preprintnumbers,amsmath,amssymb]{revtex4}
\usepackage{dcolumn}
\usepackage{bm}
\usepackage{amssymb,amsmath, amsthm,epsfig,graphicx,mathrsfs,latexsym}

\newcommand{\da}{\partial}

\begin{document}
\title{Nonlinear theory of resonant slow waves in anisotropic and dispersive plasmas}
\author{Christopher TM Clack and Istvan Ballai}
\affiliation{Department of Applied Mathematics, University of
Sheffield, Hicks Building, Hounsfield Road, Sheffield, S3 7RH, U.K.}

\begin{abstract}
The solar corona is a typical example of a plasma with strongly
anisotropic transport processes. The main dissipative mechanisms in
the solar corona acting on slow magnetoacoustic waves are the
anisotropic thermal conductivity and viscosity. Ballai \textit{et
al.} [Phys. Plasmas {\bf 5}, 252 (1998)] developed the nonlinear
theory of driven slow resonant waves in such a regime. In the
present paper the nonlinear behaviour of driven magnetohydrodynamic
waves in the slow dissipative layer in plasmas with strongly
anisotropic viscosity and thermal conductivity is expanded by
considering dispersive effects due to Hall currents. The nonlinear
governing equation describing the dynamics of nonlinear resonant
slow waves is supplemented by a term which describes
\emph{nonlinear} dispersion and is of the same order of magnitude as
nonlinearity and dissipation. The connection formulae are found to
be similar to their non-dispersive counterparts.
\end{abstract}

\maketitle

\vspace{0.1cm}

\section{Introduction}

Resonances are ubiquitous everytime magnetohydrodynamic (MHD) waves
are driven in a transversally inhomogeneous (relative to the ambient
magnetic field) plasma. From a mathematical point of view a
resonance is equivalent to regular singular points in the equations
describing the dynamics of waves, but these singularities can be
removed by, for example, dissipation. At resonance interacting
dynamical systems can transfer energy to each other. In this
context, resonant absorption has been suggested as a method to
create supplementary heating of fusion plasmas, but was, however,
later rejected due to technical difficulties (see, e.g. Refs.
\cite{tataronis1}-\cite{Hasegawa1976}). Ionson \cite{ionson1978}
suggests, for the first time, that resonant MHD waves may be a means
to heat magnetic loops in the solar corona. Since then, resonant
absorption of MHD waves has become a popular and successful
mechanism for providing some of the heating of the solar corona
(see, e.g. Refs. \cite{ofman1995}, \cite{belien1999}, and references
therein). More recently resonant absorption has acquired a new
applicability when the observed damping of waves and oscillations in
coronal loops has been attributed to resonant absorption. Hence,
resonant absorption has become a fundamental constituent block of
one of the newest branches of solar physics, called \textit{coronal
seismology} (see, e.g. Refs.
\cite{rudermanconf2002}-\cite{terradas2008}).

A driven problem for resonant MHD waves occurs when there is an
external (or internal) source of energy that excites the plasma
oscillations. If there is a small amount of dissipation present in
the system, after some time the system will attain a steady state in
which all perturbed quantities will oscillate with the same
frequency $\omega$. In a weakly dissipative plasma, dissipation only
becomes important in thin layers enclosing the singularity (the
so-called dissipative layer). Two types of driving are possible.
First, direct driving, where the source of energy is inside the
system or at the boundary and the oscillations of resonant magnetic
field lines are directly excited by this source of energy. In the
context of resonant absorption, this type of driving was studied by,
for example, Ruderman \textit{et al.} \cite{Ruderman1997a}; \cite{Ruderman1997b} and Tirry
\textit{et al.} \cite{tirry1997}. Secondly, in the case of indirect
or lateral driving, the energy source can be either outside or
inside the system. This energy source excites fast or slow
magnetosonic waves which propagate across and along magnetic
surfaces and reach the resonant magnetic surface where their energy
is partly dissipated due to resonant coupling with localized
Alfv\'{e}n or slow waves. The lateral driving problem was studied
by, for example, Davila \cite{Davila1987} for planar geometry and by
Erd\'{e}lyi \cite{erdelyi1997} for cylindrical geometry. In the
present paper, we consider only the lateral driven case (for a
comprehensive background to lateral driving see, e.g. Refs.
\cite{poedts1989}-\cite{poedts1990c}). An important property of
these waves is that their damping rate is almost independent of the
values of the dissipative coefficients, a situation characteristic
for dissipative systems with large Reynolds numbers. As a result,
the damping rate of near resonant MHD waves can be many orders of
magnitudes larger than the damping rate of MHD waves with the same
frequencies in homogeneous plasmas. The damping allows the waves
energy to be converted into heat, which has made resonant absorption
a subject of intense study.

The concept of connection formulae was introduced by Sakurai
\textit{et al.} \cite{sakurai1}, to determine the jumps in the
normal component of velocity and perturbation of total pressure
across the dissipative layer. The connection formulae avoid solving
the full dissipative MHD equations when studying resonant MHD waves.
Instead, the narrow layer embracing the resonant point can be
considered as a surface of discontinuity and ideal MHD equations can
be used to the left and right of this surface. Connection formulae
are utilized in order to connect the ideal solution over the
discontinuity.

Most studies on driven resonant MHD waves use isotropic viscosity
and/or electrical resistivity. However, the solar corona is a
well-known example of a plasma where viscosity is strongly
anisotropic \cite{Hollweg1985}. Hollweg and Yang \cite{hollweg1988b}
studied the laterally driven problem in the cold plasma
approximation. They found that anisotropic viscosity does not remove
the Alfv\'{e}n singularity (if the Braginskii's viscosity tensor is
approximated by its first term only). If Braginskii's full viscosity
tensor is considered, the Alfv\'{e}n singularity would be removed by
the shear viscosity. The way the dissipative term appears in the
governing equation is, however, identical to that of isotropic
viscosity.

For the case of slow resonant waves the situation is different. The
laterally driven linear slow resonant waves in plasmas with strongly
anisotropic viscosity and thermal conductivity was studied first by
Ruderman and Goossens \cite{Ruderman1996}. They successfully showed
that anisotropic viscosity and/or thermal conductivity removes the
slow resonance present in ideal plasmas. They also obtained the
explicit connection formulae, which are identical to those found in
the case of plasmas with isotropic viscosity and finite electrical
resistivity, \cite{sakurai1}. This fact supports the hypothesis that
in weakly dissipative plasmas the connection formulae are
independent of the exact form of dissipative processes present in
the dissipative layer.

The laterally driven nonlinear slow resonant waves in plasmas with
strongly anisotropic viscosity and thermal conductivity was first
studied by Ballai \textit{et al.} \cite{Ballai1998a}. They found
that nonlinearity was crucial in the dissipative layer. The
governing equation for slow wave dynamics in the dissipative layer
was derived and the implicit connection formulae were found,
(explicit connection formulae have only been found for linear theory
and for the limit of strong nonlinearity, \cite{Ruderman2000}). The
governing equation was almost identical to that found by Ruderman
\textit{et al.} \cite{ruderman3}, however the dissipative term was
laterally dependent ($\theta$) rather than normally dependent
($\xi$) dependent. The implicit connection formulae found coincide
with those found in plasmas with isotropic viscosity and finite
electrical resistivity \cite{ruderman3}.

A drawback of previous studies on resonant absorption is that even
though anisotropy is considered, dispersion (by, e.g. Hall effect)
is neglected. This approximation is acceptable only for lowest
regions of the solar atmosphere. The solar corona is known to be
strongly magnetized, hence the Hall term can be comparable with
other effects considered in the process of resonance. The present
paper will extend the nonlinear theory of resonant slow MHD waves in
the dissipative layer with strongly anisotropic viscosity and
thermal conductivity to include Hall dispersion and show that the
effect of this new addition is of the same order of magnitude as
nonlinearity and dissipation near resonance. The paper is organised
as follows. In the next section we introduce the fundamental
equations and discuss the main assumptions. In section III we derive
the nonlinear governing equation for waves in the dissipative layer.
Section IV is devoted to the derivation of the nonlinear analogues
of the connection formulae. Finally, in section V we summarize and
draw our conclusions pointing out a few further applications to be
carried out in the future.

\section{Fundamental Equations}

In what follows we use the visco-thermal MHD equations with strongly
anisotropic viscosity and thermal conductivity. We assume that the
plasma is strongly magnetised, so that the conditions
$\omega_e\tau_{e}\gg1$ and $\omega_i\tau_{i}\gg1$ are satisfied,
where $\omega_{e(i)}$ is the electron (ion) gyrofrequency and
$\tau_{e(i)}$ is the mean electron (ion) collision time. Due to the
strong magnetic field, transport processes are derived from
Braginskii's expression for the viscosity tensor $\hat{\pi}$ (see
Refs. \cite{Ruderman1997b} and \cite{braginskii}). Under coronal
conditions it is a good approximation to retain only the first term
of Braginskii's expression for viscosity, \cite{Hollweg1985}, namely
\begin{equation}\label{eq:braginskii}
\hat{\pi}=\bar{\eta_0}\left(\mathbf{b}\otimes\mathbf{b}-\frac{1}{3}\hat{I}\right)
\left[3\mathbf{b}\cdot(\mathbf{b}\cdot\nabla)\mathbf{v}-\nabla\cdot\mathbf{v}\right],
\end{equation}
where $\mathbf{v}$ is the velocity,
$\mathbf{b}=\overline{\mathbf{B}}/B$ is the unit vector in the
direction of magnetic field and $\bar{\eta_0}$ is the first
Braginskii coefficient of viscosity (compressional viscosity).
$\hat{I}$ is the unit tensor and the symbol $\otimes$ denotes the
dyadic product of vectors.

In a strongly magnetised plasma the thermal conductivity parallel to
the magnetic field lines dwarfs the perpendicular component so the
heat flux becomes, \cite{priest1},
\begin{equation}\label{eq:thermalconduction}
\mathbf{q}=-\bar{\kappa_{\parallel}}\mathbf{b}(\mathbf{b}\cdot\nabla\overline{T}),
\end{equation}
where $\overline{T}$ is the temperature and
$\bar{\kappa_{\parallel}}$ is the parallel coefficient of thermal
conductivity (in the solar corona,
$\overline{\kappa}_{\parallel}=9\times10^{-12}T^{5/2}\mbox{
Wm}^{-1}\mbox{K}^{-1}$).

In the solar corona the finite electrical resistivity can be
neglected as it is several orders of magnitude smaller than the
dissipative coefficients considered here, see e.g.
\cite{Erdelyi1994}.

Using Eqs. (\ref{eq:braginskii}) and (\ref{eq:thermalconduction}),
the visco-thermal MHD equations are
\begin{align}
\frac{\da\bar{\rho}}{\da
t}+\nabla\cdot(\bar{\rho}\mathbf{v})=0,\label{eq:masscontinuity}
\end{align}
\begin{multline}
\frac{\da\mathbf{v}}{\da
t}+(\mathbf{v}\cdot\nabla)\mathbf{v}=-\frac{1}{\bar{\rho}}\nabla\overline{P}+
\frac{1}{\mu\bar{\rho}}(\overline{\mathbf{B}}\cdot\nabla)\overline{\mathbf{B}}\\
+\frac{1}{\bar{\rho}}(\nabla\cdot\mathbf{b})\times(\bar{\eta}_0\mathbf{b}Q),\label{eq:momentum}
\end{multline}
\begin{equation}
\frac{\da\overline{\mathbf{B}}}{\da
t}=\nabla\times(\mathbf{v}\times\overline{\mathbf{B}})+\overline{\mathbf{H}},\label{eq:induction}
\end{equation}
\begin{multline}
\frac{\da\overline{T}}{\da
t}+\mathbf{v}\cdot\nabla\overline{T}+(\gamma-1)\overline{T}\nabla\cdot\mathbf{v}=\\
\frac{\gamma-1}{\widetilde{R}\bar{\rho}}
\left\{\nabla\cdot\left[\bar{\kappa}_{\parallel}\mathbf{b}(\mathbf{b}\cdot\nabla\overline{T})\right]
+\frac{1}{3}\bar{\eta}_0Q^{2}\right\},\label{eq:thermalconductivity}
\end{multline}
\begin{equation}
\overline{P}=\bar{p}+\frac{\overline{\mathbf{B}}^2}{2\mu}+\frac{\bar{\eta}_0}{3}Q,\label{eq:totalpressure}
\end{equation}
\begin{equation}
\bar{p}=\widetilde{R}\bar{\rho}\overline{T},\label{eq:gaslaw}
\end{equation}
\begin{equation}
\nabla\cdot\overline{\mathbf{B}}=0,\label{eq:solenoid}
\end{equation}
\begin{equation}
Q=3\mathbf{b}\cdot(\mathbf{b}\cdot\nabla)\mathbf{v}-\nabla\cdot\mathbf{v}.\label{eq:Q}
\end{equation}
In Eqs. (\ref{eq:masscontinuity})-(\ref{eq:Q}), $\bar{p}$ is the
kinematic pressure, $\bar{\rho}$ the density, $\overline{P}$ the
viscosity modified total pressure (kinetic and magnetic), $\gamma$
the adiabatic exponent, $\tilde{R}$ the gas constant and $\mu$ the
magnetic permeability. The term $\overline{\mathbf{H}}$ in Eq.
(\ref{eq:induction}) is the Hall term given by
\begin{equation}\label{eq:hallterm}
\overline{\mathbf{H}}=\frac{1}{\mu e}\nabla\times
\left(\frac{1}{n_e}\overline{\mathbf{B}}\times\nabla\times\overline{\mathbf{B}}\right),
\end{equation}
where $e$ is the electron charge and $n_e$ is the electron number
density. The propagation of compressional linear and nonlinear MHD
waves in Hall plasmas has been studied by, for example, Baronov \&
Ruderman \cite{baronov1974}, Ruderman \cite{ruderman1976};
\cite{ruderman1987}; \cite{ruderman2002}, Ballai \textit{et al.}
\cite{ballai2003} and Miteva \textit{et al.} \cite{miteva2004}. As
stated in an earlier study by Huba \cite{huba1995}, Hall MHD is only
relevant to plasma dynamics occurring on length scales of the order
of the ion inertial length, $d_i=c/\omega_i$, where $c$ is the speed
of light and $\omega_i$ is the ion plasma frequency. For the present
paper this would require that $d_i=\mathscr{O}(\delta_c)$, where
$\delta_c$ is the thickness of the dissipative layer. Indeed,
starting from the upper chromosphere this condition is satisfied and
the lengths involved in the problem are of the order of $10-100$m.
Hall currents arise when considering off diagonal terms in the
conductivity tensor.

We adopt a Cartesian coordinate system, and limit our analysis to a
static background equilibrium ($\mathbf{v}_0=0$). We assume that all
equilibrium quantities depend on $x$ only. The equilibrium magnetic
field, $\mathbf{B}_0$, is unidirectional and lies in the $yz$-plane.
The equilibrium quantities must satisfy the condition of total
pressure balance,
\begin{equation}\label{eq:pressurebalance}
p_0+\frac{B_{0}^2}{2\mu}=constant.
\end{equation}
In addition we assume that the wave propagation is independent of
$y$ ($\da/\da y =0$). In linear theory of driven waves all perturbed
quantities oscillate with the same frequency $\omega$, so they can
be Fourier-analysed and taken to be proportional to
$\exp(i[kz-\omega t])$, so solutions are sought in the form of
propagating waves. All perturbations in these solutions depend on
the combination $\theta=z-Vt$, rather than $z$ and $t$ separately,
with $V=\omega/k$. In the context of resonant absorption the phase
velocity, $V$, must match the projection of the cusp velocity,
$c_T$, onto the $z$-axis when $x=x_c$. To define the resonant
position mathematically it is convenient to introduce the angle,
$\alpha$, between the $z$-axis and the direction of the equilibrium
magnetic field, so that the components of the equilibrium magnetic
field are
\begin{equation}\label{eq:angles}
B_{0y}=B_0\sin\alpha,\phantom{X}B_{0z}=B_0\cos\alpha.
\end{equation}
The definition of the resonant position can now be written mathematically as
\begin{equation}\label{eq:resonantposition}
V=c_{T}\left(x_c\right)\cos\alpha.
\end{equation}
The square of the cusp
speed is defined as
\begin{equation}\label{eq:cuspspeed}
c_T^2=\frac{c_S^2v_A^2}{c_S^2+v_A^2},
\end{equation}
where the squares of the sound and Alfv\'{e}n speeds are given by
\begin{equation}\label{eq:alfvensound}
c_S^2=\frac{\gamma p_0}{\rho_0},\phantom{X}
v_A^2=\frac{B_0^2}{\mu\rho_0},
\end{equation}
with all speeds being dependent on the coordinate $x$.

In a nonlinear regime the perturbations cannot Fourier-analysed,
however, in an attempt to adhere as close to linear theory as
possible we look for travelling wave solutions and assume all
perturbed quantities depend on $\theta=z-Vt$ where $V$ is given by
Eq. (\ref{eq:resonantposition}). The perturbations of the physical
quantities are defined by
\begin{align}
&\bar{\rho}=\rho_0+\rho,\phantom{X} \bar{p}=p_0+p,\phantom{X}
\overline{T}=T_0+T\nonumber\\
&\overline{\mathbf{B}}=\mathbf{B}_0+\mathbf{B},\phantom{X}
\overline{\mathbf{H}}=\mathbf{H}_0+\mathbf{H},\phantom{X}\overline{P}=P_0+\widetilde{P}\label{eq:perturbations}
\end{align}
It is clear from Equations (\ref{eq:hallterm}) and
(\ref{eq:perturbations}) that $\mathbf{H}_0=0$.

The dominant dynamics of resonant slow waves, in linear MHD, resides
in the components of magnetic field and velocity that are parallel
to the equilibrium magnetic field (as well as in the compressional
quantities $\rho$, $p$ and $T$). This dominant behaviour is created
by an $x^{-1}$ singularity in the spatial solution of these
quantities at the cusp resonance, (\cite{sakurai1}); these are known
as \emph{large variables}. The normal components of velocity, $u$,
and magnetic field, $B_x$, are also singular however their
singularity is proportional to $\ln|x|$. In addition, the quantities
$\overline{P}$ and the components of $\mathbf{B}$ and $\mathbf{v}$
that are perpendicular to the equilibrium magnetic field are
regular; these are known as \emph{small variables}.

To make the mathematical analysis more concise and to make the
physics more transparent we define the components of velocity and
magnetic field that are parallel and perpendicular to the
equilibrium magnetic field (as well as existing in the $yz$-plane):
\begin{align}
&\left(
\begin{array}{cc}
v_{\parallel}\\
B_{\parallel}
\end{array}\right)=\left(
\begin{array}{cc}
v\phantom{X} w\\
B_y\phantom{X} B_z
\end{array}\right)
\left(\begin{array}{cc}
\sin\alpha\\
\cos\alpha
\end{array}\right),\nonumber\\
&\left(
\begin{array}{cc}
v_{\perp}\\
B_{\perp}
\end{array}\right)=\left(
\begin{array}{cc}
v\phantom{X} -w\\
B_y\phantom{X} -B_z
\end{array}\right)
\left(\begin{array}{cc}
\cos\alpha\\
\sin\alpha
\end{array}\right).
\end{align}
where $u$, $w$, $B_y$ and $B_z$ are the $y$- and $z$-components of
the velocity and perturbation of magnetic field, respectively.

Let us introduce the characteristic scale of inhomogeneity,
$l_{inh}$. The classic Reynolds number, $R_e$, and the Pecklet
number, $P_e$, are defined as
\begin{equation}\label{eq:reynoldsnumbers}
R_e=\frac{\rho_{0_c}Vl_{inh}}{\bar{\eta}_{0_c}},\phantom{x}P_e=\frac{\rho_{0_c}\widetilde{R}Vl_{inh}}{\bar{\kappa}_{\parallel_c}}.
\end{equation}
These two numbers determine the importance of viscosity and thermal
conduction. We introduce the total Reynolds number as
\begin{equation}\label{eq:totalreynolds}
\frac{1}{R}=\frac{1}{R_e}+\frac{1}{P_e}.
\end{equation}

The aim of this paper is to derive the governing equation for waves
in the slow dissipative layers taking into account nonlinearity,
dissipation and dispersion simultaneously. Since for the coronal
plasma $R\gg1$, we consider only weakly dissipative plasmas. We
introduce, $\epsilon$, as the dimensionless amplitude of variables
far away from the resonance. Linear theory predicts that the
characteristic scale of dissipation, $l_{diss}$, is of the order
$R^{-1}l_{inh}$. The typical largest nonlinear term in the system of
MHD equations is of the form $g\da g/\da z$ while the typical
dissipative term if of the form $\bar{\eta}_0\da^2 g/\da z^2$, where
$g$ is any `large' variable. Linear theory shows that `large'
variables have an ideal singularity $(x-x_c)^{-1}$ in the vicinity
of $x=x_c$. This implies that the `large' variables have
dimensionless amplitudes, inside the dissipative layer, of the order
of $\epsilon R$. It is now straightforward to estimate the ratio of
a typical nonlinear and dissipative term,
\begin{equation}\label{eq:ratioofnonlindiss}
\phi=\frac{g\da g/\da z}{\bar{\eta}_0\da^2 g/\da
z^2}=\mathscr{O}(\epsilon R^{2}),
\end{equation}
where the quantity $\phi$ can be considered as a nonlinearity
parameter. If the condition $\epsilon R^2\ll1$ is satisfied, linear
theory is applicable. On the other hand, if $\epsilon R^2\gtrsim1$
then nonlinearity has to be taken into account when studying
resonant waves in the dissipative layers. Using the same scalings,
Ballai \textit{et al.} \cite{Ballai1998a} showed that nonlinearity
has to be considered whenever slow wave resonant absorption is
studied in the solar corona. This conclusion simply means that in
the solar upper atmosphere resonant absorption is a nonlinear
phenomena.

In order to obtain nonlinearity and dissipation of equal order we
assume that $\epsilon R^2=\mathscr{O}(1)$, i.e. $R\sim
\epsilon^{-1/2}$, when deriving the nonlinear governing equations
for waves in the slow dissipative layer. Accordingly, we can
re-scale the coefficients of viscosity and thermal conductivity as
\begin{equation}\label{eq:rescalecoeffs}
\bar{\eta}_0=\epsilon^{1/2}\eta_0,\phantom{X}\bar{\kappa}_{\parallel}
=\epsilon^{1/2}\kappa_{\parallel}.
\end{equation}
We also consider the coefficient of Hall conduction, defined as
$\overline{\chi}=\overline{\eta}\omega_e\tau_e$, where
$\overline{\eta}$ is the coefficient of magnetic diffusivity
(although $\overline{\eta}$ is small enough, in the solar corona, to
be neglected in comparison to $\overline{\eta}_0$, here it is
multiplied by the product $\omega_e\tau_e$ which is very large under
coronal conditions). Similar to the previous dissipative
coefficients, we introduce the scaling
\begin{equation}
\overline{\chi}=\epsilon^{1/2}\chi.
\end{equation}
Using Eq. (\ref{eq:rescalecoeffs}), we can rewrite Eqs.
(\ref{eq:masscontinuity})-(\ref{eq:totalpressure}) in the form
\begin{multline}\label{eq:masscontinuity1}
V\frac{\da \rho}{\da\theta}-\frac{\da(\rho_0 u)}{\da
x}-\rho_0\frac{\da w}{\da\theta}=\frac{\da(\rho u)}{\da
x}+\frac{\da(\rho w)}{\da\theta},
\end{multline}
\begin{multline}\label{eq:momentumx1}
\rho_0 V\frac{\da u}{\da\theta}-\frac{\da\widetilde{P}}{\da
x}+\frac{B_0 \cos\alpha}{\mu}\frac{\da B_x}{\da\theta}=
\bar{\rho}\left(u\frac{\da u}{\da x}+w\frac{\da
w}{\da\theta}\right)\\
-\rho V\frac{\da u}{\da\theta}-\frac{B_x}{\mu}\frac{\da
B_x}{\da x}-\frac{B_z}{\mu}\frac{\da B_x}{\da\theta}\\
-\epsilon^{1/2}\left(\frac{\da}{\da
x}b_x+\frac{\da}{\da\theta}b_z\right)\left(\eta_0 b_{x}Q\right),
\end{multline}
\begin{multline}\label{eq:momentumperp1}
\frac{\da}{\da\theta}\left(\rho_0 V
v_{\perp}+\widetilde{P}\sin\alpha+\frac{B_0\cos\alpha}{\mu}B_{\perp}\right)=\\
\bar{\rho}\left(u\frac{\da v_{\perp}}{\da x}+w\frac{\da
v_{\perp}}{\da\theta}\right) -\rho V\frac{\da
v_{\perp}}{\da\theta}-\frac{B_x}{\mu}\frac{\da
B_{\perp}}{\da x}-\frac{B_z}{\mu}\frac{\da B_{\perp}}{\da\theta}\\
-\epsilon^{1/2}\left(\frac{\da}{\da
x}b_x+\frac{\da}{\da\theta}b_z\right)\left(\eta_0 b_{\perp}Q\right),
\end{multline}
\begin{multline}\label{eq:momentumpara1}
\frac{\da}{\da\theta}\left(\rho_0 V
v_{\parallel}-\widetilde{P}\cos\alpha+\frac{B_0\cos\alpha}{\mu}B_{\parallel}\right)+\frac{B_x}{\mu}\frac{dB_0}{dx}=\\
\bar{\rho}\left(u\frac{\da v_{\parallel}}{\da x}+w\frac{\da
v_{\parallel}}{\da\theta}\right) -\rho V\frac{\da
v_{\parallel}}{\da\theta}-\frac{B_x}{\mu}\frac{\da
B_{\parallel}}{\da x}-\frac{B_z}{\mu}\frac{\da B_{\parallel}}{\da\theta}\\
-\epsilon^{1/2}\left(\frac{\da}{\da
x}b_x+\frac{\da}{\da\theta}b_z\right)\left(\eta_0
b_{\parallel}Q\right),
\end{multline}
\begin{multline}\label{eq:inductx1}
VB_x+B_0 u\cos\alpha=wB_x-uB_z\\
-\epsilon^{1/2}\chi\frac{\da
B_{\parallel}}{\da\theta}\cos\alpha\sin\alpha,
\end{multline}
\begin{multline}\label{eq:inductperp1}
\frac{\da}{\da\theta}\left(VB_{\perp}+B_0
v_{\perp}\cos\alpha\right)=\frac{\da(uB_{\perp})}{\da
x}+\frac{\da(wB_{\perp})}{\da\theta}\\
-B_x\frac{\da v_\perp}{\da x}-B_z\frac{\da v_{\perp}}{\da\theta}
-\epsilon^{1/2}\chi\frac{\da^2 B_{\parallel}}{\da
x\da\theta}\cos\alpha,
\end{multline}
\begin{multline}\label{eq:inductpara1}
\frac{\da}{\da\theta}\left(VB_{\parallel}+B_0
v_{\parallel}\cos\alpha\right)-\frac{\da(B_0 u)}{\da
x}-B_0\frac{\da w}{\da\theta}=\\
\frac{\da(uB_{\parallel})}{\da
x}+\frac{\da(wB_{\parallel})}{\da\theta}-B_x\frac{\da
v_{\parallel}}{\da x}-B_z\frac{\da v_{\parallel}}{\da\theta}\\
-\epsilon^{1/2}\frac{\chi}{\rho_0}\frac{\da
B_{\parallel}}{\da\theta}\frac{\da\rho}{\da x}\sin\alpha,
\end{multline}
\begin{equation}\label{eq:solenoid1}
\frac{\da B_x}{\da x}+\frac{\da B_z}{\da\theta}=0,
\end{equation}
\begin{multline}\label{eq:thermal1}
V\frac{\da
T}{\da\theta}-u\frac{dT_0}{dx}-(\gamma-1)T_0\left(\frac{\da u}{\da
x}+\frac{\da w}{\da\theta}\right)=\\
u\frac{\da T}{\da x}+w\frac{\da
T}{\da\theta}+(\gamma-1)T\left(\frac{\da u}{\da x}+\frac{\da
w}{\da\theta}\right)\\
-\epsilon^{1/2}\frac{\gamma-1}{\bar{\rho}\widetilde{R}}\left\{\frac{1}{3}\eta_0Q^2+\left(\frac{\da}{\da
x}b_x+\frac{\da}{\da\theta}b_z\right)\right.\\
\left.\times\kappa_{\parallel}\left[b_x\left(\frac{dT_0}{dx}+\frac{\da
T}{\da x}\right)+b_z\frac{\da T}{\da\theta}\right]\right\},
\end{multline}
\begin{multline}\label{eq:totalpressure1}
\widetilde{P}=p+\frac{1}{2\mu}\left(B_x^2+B_{\perp}^2+B_{\parallel}^2+2B_0
B_{\parallel}\right)+\frac{1}{3}\epsilon^{1/2}\eta_0 Q,
\end{multline}
\begin{equation}\label{eq:gas1}
\frac{\gamma T_0 p}{c_S^2}-T_0\rho-\rho_0 T=\rho T,
\end{equation}
\begin{multline}\label{eq:Q1}
Q=3b_x\left(b_x\frac{\da u}{\da x}+b_z\frac{\da
u}{\da\theta}\right)+3b_{\parallel}\left(b_x\frac{\da
v_{\parallel}}{\da x}+b_z\frac{\da
v_{\parallel}}{\da\theta}\right)\\
-\left(\frac{\da u}{\da x}+\frac{\da w}{\da\theta}\right).
\end{multline}
We should state that in Eqs.
(\ref{eq:inductx1})-(\ref{eq:inductpara1}) we have used the
coefficient of Hall conduction, $\chi$, which does not contribute to
the total Reynolds number because it is the multiplier of dispersive
terms rather than dissipative ones. The derivation of the
expressions of the Hall terms in the induction equation can be found
in the Appendix.

The set of Eqs. (\ref{eq:masscontinuity1})-(\ref{eq:Q1}), will be
used in the next section to derive the governing equation for wave
motion in the dissipative layer.

\section{Deriving the Governing Equation in the Dissipative Layer}

In order to derive the governing equation for wave motions in the
slow dissipative layer we employ the method of matched asymptotic
expansions (\cite{nayfeh1981}). This method requires us to find the
so-called \textit{outer} and \textit{inner} expansions and then
match them in the overlap regions. This nomenclature is ideal for
our situation. The outer expansion corresponds to the solution
outside the dissipative layer and the inner expansion corresponds to
the solution inside the dissipative layer. A simplified version of
the method of matched asymptotic expansions, developed by Ballai
\textit{et al.} \cite{Ballai1998a}, is adopted here.

We only consider weakly dissipative plasmas, so viscosity and
thermal conductivity are only important in the narrow dissipative
layer (here dissipation and nonlinearity are of the same order)
embracing the resonant position. Far away from the dissipative layer
the amplitudes of perturbations are small, so we use linear ideal
MHD equations in order to describe the wave motion far away from the
dissipative layer. The full set of nonlinear dissipative MHD
equations are used for describing wave motion \textit{inside} the
dissipative layer where the amplitudes are much larger than those
far away from the dissipative layer. We therefore look for solutions
in the form of asymptotic expansions. The equilibrium quantities
change only slightly across the dissipative layer so it is possible
to approximate them by the first non-vanishing term in their Taylor
series expansion with respect to $x$. Similar to linear theory, we
assume the expansions of equilibrium quantities are valid in a
region embracing the ideal resonant position, which is assumed to be
much wider than the dissipative layer, $l_{inh}/l_{diss}\sim R$.
This implies there are two overlap regions, one to the left and one
to the right of the dissipative layer, where both the outer (the
solution to the linear ideal MHD equations) and inner (the solution
to the nonlinear dissipative MHD equations) solutions are valid.
Therefore, both solutions must coincide in the overlap regions which
provides the matching conditions.

Before deriving the nonlinear governing equation we ought to make a
note. In linear theory, perturbations of physical quantities are
harmonic functions of $\theta$ and their mean values vanish over a
period. In nonlinear theory, however, the perturbations of variables
can have non-zero values as a result of nonlinear interaction of
different harmonics. Due to the nonlinear absorption of wave
momentum, a mean shear flow is generated outside the dissipative
layer, as shown by Ruderman \textit{et al.} \cite{ruderman3} in
Cartesian geometry. In our scenario a mean shear flow is created
outside the dissipative layer, but as there is no perpendicular
component to viscosity oscillating plasma can slide past each other
without friction. This produces a mean flow with infinite amplitude.
However, boundaries can prevent such oscillations. Therefore, we
will assume such boundaries exist and will not consider the
generation of mean shear flow.

The first step in our mathematical description is the derivation of
governing equations outside the dissipative layer where the dynamics is
described by ideal ($\eta_0=\kappa_{\parallel}=0$) and linear MHD. Since dispersion is assumed to act over scales comparable
to dissipative scales, dispersion will only be taken into account inside the dissipative layer.
The linear form of Eqs. (\ref{eq:masscontinuity1})-(\ref{eq:Q1}) can
be obtained by assuming a regular expansion of variables of the form
\begin{equation}\label{eq:linearexpansion}
f=\epsilon f^{(1)}+\epsilon^{3/2} f^{(2)}\ldots,
\end{equation}
and collect only terms proportional to the small parameter
$\epsilon$. This leads to a system of linear equations for the
variables with superscript `1'. All variables can be eliminated,
with the exception of $u^{(1)}$ and $P^{(1)}$, leading to the system
\begin{equation}\label{eq:linearequation}
\frac{\da u^{(1)}}{\da x}=\frac{V}{F}\frac{\da P^{(1)}}{\da\theta},\phantom{X}
\frac{\da P^{(1)}}{\da x}=\frac{\rho_0 A}{V}\frac{\da u^{(1)}}{\da\theta},
\end{equation}
where
\begin{equation}\label{eq:F}
F=\frac{\rho_0 A C}{V^4-V^2\left(v_A^2+c_S^2\right)+v_A^2
c_S^2\cos^{2}\alpha},
\end{equation}
\begin{equation}\label{eq:A}
A=V^2-v_A^2\cos^{2}\alpha,\nonumber
\end{equation}
\begin{equation}\label{eq:C}
C=\left(v_A^2+c_S^2\right)\left(V^2-c_T^2\cos^{2}\alpha\right).
\end{equation}
The reason for using $P$ instead of $\widetilde{P}$ is that outside
the dissipative layer the plasma is ideal, so there is no viscous
addition to the total pressure. The remaining variables can all be
expressed in terms of $u^{(1)}$ and $P^{(1)}$,
\begin{equation}\label{eq:relationlinear1}
v_{\perp}^{(1)}=-\frac{V\sin\alpha}{\rho_0 A}P^{(1)}, \phantom{X}
v_{\parallel}^{(1)}=\frac{Vc_S^2\cos\alpha}{\rho_0 C}P^{(1)},
\end{equation}
\begin{equation}\label{eq:relationlinear2}
B_x^{(1)}=-\frac{B_0\cos\alpha}{V}u^{(1)},\phantom{X}
B_{\perp}^{(1)}=\frac{B_0\cos\alpha\sin\alpha}{\rho_0 A}P^{(1)},
\end{equation}
\begin{equation}\label{eq:relationlinear3}
\frac{\da B_{\parallel}^{(1)}}{\da\theta}=\frac{B_0\left(
V^2-c_S^2\cos^{2}\alpha\right)}{\rho_0 C}\frac{\da
P^{(1)}}{\da\theta} +\frac{u^{(1)}}{V}\frac{dB_0}{dx},
\end{equation}
\begin{equation}\label{eq:relationlinear4}
\frac{\da p^{(1)}}{\da\theta}=\frac{V^2 c_S^2}{C}
\frac{\da P^{(1)}}{\da\theta}-\frac{u^{(1)}B_0}{\mu V}\frac{dB_0}{dx},
\end{equation}
\begin{equation}\label{eq:relationlinear5}
\frac{\da \rho^{(1)}}{\da\theta}=\frac{V^2}{C}
\frac{\da P^{(1)}}{\da\theta}+\frac{u^{(1)}}{V}\frac{d\rho_0}{dx},
\end{equation}
\begin{equation}\label{eq:relationlinear6}
\frac{\da T^{(1)}}{\da\theta}=\frac{(\gamma-1)T_0 V^2}{\rho_0 C}
\frac{\da P^{(1)}}{\da\theta}+\frac{u^{(1)}}{\gamma V\widetilde{R}}
\frac{dc_S^2}{dx}.
\end{equation}
The differential Eq. (\ref{eq:linearequation}) have regular
singularities at the resonance, therefore the solutions can be
obtained in terms of Fr\"{o}benius series with respect to $x$ (for
details see, e.g. \cite{Ballai1998a}; \cite{ruderman3}) of the form
\begin{equation}\label{eq:frobenius1}
P^{(1)}=P_1^{(1)}(\theta)+P_2^{(1)}(\theta)x\ln|x|+P_3^{(1)}(\theta)+\ldots,
\end{equation}
\begin{equation}\label{eq:frobenius2}
u^{(1)}=u_1^{(1)}(\theta)\ln|x|+u_2^{(1)}(\theta)+u_3^{(1)}(\theta)x\ln|x|+\ldots.
\end{equation}
The coefficient functions depending on $\theta$ in the above expansions are
generally different for $x<0$ and $x>0$. The salient property of
these solutions is that the perturbation of the total pressure is
regular at the ideal resonant position $x=x_c$. From Eqs.
(\ref{eq:relationlinear1})-(\ref{eq:relationlinear6}), we see that
the quantities $v_{\perp}^{(1)}$ and $B_{\perp}^{(1)}$ are also
regular at $x=x_c$. All other quantities are singular. The
quantities $u^{(1)}$ and $B_x^{(1)}$ behave as $\ln|x|$, while
$v_{\parallel}^{(1)}$, $B_{\parallel}^{(1)}$, $p^{(1)}$,
$\rho^{(1)}$ and $T^{(1)}$ behave as $x^{-1}$, so they are the most
singular. This property can be extended and will hold to all higher
order approximations, \cite{ruderman3}.

Now let us concentrate on the solution in the dissipative layer. The
thickness of the dissipative layer is of the order $l_{inh}R^{-1}$.
We have assumed that $R\sim \epsilon^{-1/2}$ so we obtain
$l_{inh}R^{-1}=\mathscr{O}(\epsilon^{1/2}l_{inh})$. The implication
of this scaling is that we have to introduce a new stretched variable to
replace the transversal coordinate in the dissipative layer, so we
are going to use $\xi=\epsilon^{-1/2}x$. Equations
(\ref{eq:masscontinuity1})-(\ref{eq:Q1}) are not rewritten here as
they are easily obtained by the substitution of
\begin{equation}\label{eq:subs}
\frac{\da}{\da x}=\epsilon^{-1/2}\frac{\da}{\da\xi},
\end{equation}
for all derivatives. The equilibrium quantities still depend on $x$,
not $\xi$ (their expression is valid in a wider region than the
characteristic scale of dissipation). All equilibrium quantities are
expanded around the ideal resonant position, $x=x_c$, as
\begin{multline}\label{eq:equilibsub}
f_0=f_{0_c}+\xi\left(\frac{\da f_0}{\da\xi}\right)_c+\ldots\\
\approx f_{0_c}+\epsilon^{1/2} \xi\left(\frac{df_0}{dx}\right)_c,
\end{multline}
where $f_0$ is any equilibrium quantity and the subscript `c'
indicates the equilibrium quantity has been evaluated at the
resonant point (we can always make $x_c=0$ by proper translation of
the coordinate system).

We seek the solution to the set of equations obtained from Eqs.
(\ref{eq:masscontinuity1})-(\ref{eq:Q1}) by the substitution of
$x=\epsilon^{1/2}\xi$ into variables in the form of power series of $\epsilon$.
These equations contain powers of $\epsilon^{1/2}$, so we use this
quantity as an expansion parameter. To derive the form of the
inner expansions of different quantities we have to analyze the
outer solutions. First, since $v_{\perp}$ and $B_{\perp}$ are regular
at $x=x_c$ we can write their inner expansions in the form of their outer
expansions, namely Eq. (\ref{eq:linearexpansion}). The quantity
$\widetilde{P}$ is the sum of the perturbation of total pressure
$P$, which is regular at $x=x_c$, and the dissipative term
proportional to $Q$. From Eq. (\ref{eq:Q1}) it is obvious that $Q$
behaves as $x^{-1}$ in the vicinity of $x=x_c$. Far away from the
dissipative layer, $Q$ is of the order $\epsilon$. Since the
thickness of the dissipative layer is of the order
$\epsilon^{1/2}l_{inh}$, $Q$ is of the order $\epsilon^{1/2}$ in the
dissipative layer. However, Eq. (\ref{eq:totalpressure1}) clearly
shows the term proportional to $Q$ contains a multiplier,
$\epsilon^{1/2}$, which implies the contribution of $\widetilde{P}$
supplied by the dissipative term is of the order $\epsilon$. As a
consequence, we write the inner expansion of $\widetilde{P}$ in the
form of its outer expansion, Eq. (\ref{eq:linearexpansion}). The
amplitudes of large variables in the dissipative layer are of the
order $\epsilon^{1/2}$, so the inner expansion of the variables
$v_{\parallel}$, $B_{\parallel}$, $p$, $\rho$ and $T$ is
\begin{equation}\label{eq:innerexpansion}
g=\epsilon^{1/2}g^{(1)}+\epsilon g^{(2)}+\ldots.
\end{equation}
The quantities $u$ and $B_x$ behave as $\ln|x|$ in the vicinity of
$x=x_c$, which suggests that they have expansions with terms of the
order of $\epsilon\ln\epsilon$ in the dissipative layer. Ruderman
\textit{et al.} \cite{ruderman3} showed that, strictly speaking, the
inner expansions of all variables have to contain terms proportional
to $\epsilon\ln\epsilon$ and $\epsilon^{3/2}\ln\epsilon$. In the
simplified version of matched asymptotic expansions,
\cite{Ballai1998a}, we utilize the fact that
$|\ln\epsilon|\ll\epsilon^{-\kappa}$ for any positive $\kappa$ and
$\epsilon\rightarrow +0$, and consider $\ln\epsilon$ as a quantity
of the order of unity. This enables us to write the inner expansions
for $u$ and $B_x$ in the form of Eq. (\ref{eq:linearexpansion}).

We now substitute the expansion (\ref{eq:linearexpansion}) for $u$,
$B_x$, $\widetilde{P}$, $v_{\perp}$, $B_{\perp}$ and the expansion
given by (\ref{eq:innerexpansion}) for $v_{\parallel}$,
$B_{\parallel}$, $p$, $\rho$, $T$ into the set of equations obtained
from Eqs. (\ref{eq:masscontinuity1})-(\ref{eq:totalpressure1}) after
substitution of $x=\epsilon^{1/2}\xi$. The first order approximation
(terms proportional to $\epsilon$), yields a linear homogeneous
system of equations for the terms with superscript `1'. The
important result that follows from this set of equations is that
\begin{equation}\label{eq:pressuretheta}
\widetilde{P}^{(1)}=\widetilde{P}^{(1)}(\theta),
\end{equation}
that is to say $\widetilde{P}^{(1)}$ does not change across the
dissipative layer. This result parallels the result found in linear
theory (\cite{sakurai1}; \cite{goossens1995}) and nonlinear theory
(\cite{Ballai1998a}; \cite{ruderman3}). Subsequently, all remaining
variables can be expressed in terms of $u^{(1)}$,
$v_{\parallel}^{(1)}$ and $\widetilde{P}^{(1)}$ as
\begin{align}
& v_{\perp}^{(1)}=\frac{c_{S_c}^2\sin\alpha}{\rho_{0_c} V
v_{A_c}^2}\widetilde{P}^{(1)}(\theta),
\phantom{x}B_x^{(1)}=-\frac{B_{0_c}\cos\alpha}{V}u^{(1)},\label{eq:relations1}\\
& B_{\perp}^{(1)}=-\frac{B_{0_c}c_{S_c}^2\sin\alpha\cos\alpha}
{\rho_{0_c}V^2v_{A_c}^2}\widetilde{P}^{(1)}(\theta),\label{eq:relations2}\\
&B_{\parallel}^{(1)}=-\frac{B_{0_c}V}{v_{A_c}^2\cos\alpha}v_{\parallel}^{(1)},
\phantom{x}p^{(1)}=\frac{\rho_{0_c}V}{\cos\alpha}v_{\parallel}^{(1)},\label{eq:relations3}\\
&\rho^{(1)}=\frac{\rho_{0_c}V}{c_{S_c}^2\cos\alpha}v_{\parallel}^{(1)},
\phantom{x}T^{(1)}=\frac{(\gamma-1)T_{0_c}V}{c_{S_c}^2\cos\alpha}v_{\parallel}^{(1)}.\label{eq:relations4}
\end{align}
In addition, we find that the equation that relates $u^{(1)}$ and
$v_{\parallel}^{(1)}$ is
\begin{align}
&\frac{\da u^{(1)}}{\da\xi}+\frac{V^2}{v_{A_c}^2\cos\alpha}\frac{\da
v_{\parallel}^{(1)}}{\da\theta}=0.\label{eq:relations5}
\end{align}

In the second order approximation we use only the expressions
obtained from Eqs. (\ref{eq:masscontinuity1}),
(\ref{eq:momentumpara1}), (\ref{eq:inductpara1}),
(\ref{eq:thermal1}), (\ref{eq:gas1}) and (\ref{eq:totalpressure1}).
Employing Eqs. (\ref{eq:pressuretheta})-(\ref{eq:relations5}), we replace the variables in the first order approximation.
The equations obtained in the second order approximation are
\begin{multline}\label{eq:secordconserve}
\rho_{0_c}\left(\frac{\da u^{(2)}}{\da\xi}+\frac{\da
v_{\parallel}^{(2)}}{\da\theta}\cos\alpha\right)-V\frac{\da\rho^{(2)}}{\da\theta}=\\
-u^{(1)}\left(\frac{d\rho_0}{dx}\right)_c-\frac{v_{A_c}^2\cos\alpha}{v_{A_c}^2+c_{S_c}^2}
\left(\frac{d\rho_0}{dx}\right)_c\xi\frac{\da
v_{\parallel}^{(1)}}{\da\theta}\\
+\frac{c_{S_c}^2\sin^2\alpha}{Vv_{A_c}^2}\frac{d\widetilde{P}}{d\theta}
-\frac{\rho_{0_c}V}{c_{S_c}^2}\left(\frac{u^{(1)}}{\cos\alpha}\frac{\da
v_{\parallel}^{(1)}}{\da\xi}\right.\\
\left.+\frac{2v_{A_c}^2
+c_{S_c}^2}{c_{S_c}^2+v_{A_c}^2}v_{\parallel}^{(1)}\frac{\da
v_{\parallel}^{(1)}}{\da\theta}\right),
\end{multline}
\begin{multline}\label{eq:secordparamoment}
\frac{\da}{\da\theta}\left(Vv_{\parallel}^{(2)}+\frac{B_{0_c}\cos\alpha}{\mu\rho_{0_c}}B_{\parallel}^{(2)}\right)
=\frac{\cos\alpha}{\rho_{0_c}}\frac{d\widetilde{P}^{(1)}}{d\theta}\\
+\frac{B_{0_c}\cos\alpha}{\mu
v\rho_{0_c}}u^{(1)}\left(\frac{dB_0}{dx}\right)_c+\frac{V}{B_{0_c}}\left[\left(\frac{dB_0}{dx}\right)_c
\right.\\
\left.-\frac{B_{0_c}}{\rho_{0_c}}\left(\frac{d\rho_0}{dx}\right)_c\right]-\frac{\eta_0\cos^2\alpha}{\rho_{0_c}}
\frac{2v_{A_c}^2+3c_{S_c}^2}{v_{A_c}^2+c_{S_c}^2}\frac{\da^2
v_{\parallel}^{(1)}}{\da\theta^2},
\end{multline}
\begin{multline}\label{eq:secordparainduct}
V\frac{\da B_{\parallel}^{(2)}}{\da\theta}-B_{0_c}\frac{\da
u^{(2)}}{\da\xi}=u^{(1)}\left(\frac{dB_0}{dx}\right)_c\\
-\frac{V^2}{v_{A_c}^2\cos\alpha}\left(\frac{dB_0}{dx}\right)_c\xi\frac{\da
v_{\parallel}^{(1)}}{\da\theta}-\frac{B_{0_c}c_{S_c}^2\sin^2\alpha}{\rho_{0_c}Vv_{A_c}^2}
\frac{d\widetilde{P}^{(1)}}{d\theta}\\
+\frac{B_{0_c}v_{A_c}^2}{V\left(v_{A_c}^2+c_{S_c}^2\right)}\left(u^{(1)}\frac{\da
v_{\parallel}^{(1)}}{\da\xi}\cos\alpha-\frac{V^2}{v_{A_c}^2}v_{\parallel}^{(1)}\frac{\da
v_{\parallel}^{(1)}}{\da\theta}\right)\\
+\chi\frac{B_{0_c}\sin\alpha}{v_{A_c}^2+c_{S_c}^2}\frac{\da
v_{\parallel}^{(1)}}{\da\xi}\frac{\da
v_{\parallel}^{(1)}}{\da\theta},
\end{multline}
\begin{multline}\label{eq:secordthermal}
V\frac{\da T^{(2)}}{\da\theta}-(\gamma-1)T_{0_c}\left(\frac{\da
u^{(2)}}{\da\xi}+\frac{\da
v_{\parallel}}{\da\theta}\cos\alpha\right)=\\
u^{(1)}\left(\frac{dT_0}{dx}\right)_c
+(\gamma-1)\left[\frac{v_{A_c}^2\cos\alpha}{T_{0_c}\left(v_{A_c}^2+c_{S_c}^2\right)}
\xi\left(\frac{dT_0}{dx}\right)_c\frac{\da
v_{\parallel}^{(1)}}{\da\theta}\right.\\
\left.-\frac{c_{S_c}^2\sin^2\alpha}{\rho_{0_c}Vv_{A_c}^2}\frac{d\widetilde{P}}{d\theta}
-\frac{(\gamma-1)V\kappa_{\parallel}\cos\alpha}{\rho_{0_c}\widetilde{R}c_{S_c}^2}\frac{\da^2
v_{\parallel}}{\da\theta^2}\right.\\
\left.+\frac{V}{c_{S_c}^2}\left(\frac{u^{(1)}}{\cos\alpha}\frac{\da
v_{\parallel}^{(1)}}{\da\xi}+\frac{\gamma
v_{A_c}^2+c_{S_c}^2}{v_{A_c}^2+c_{S_c}^2}v_{\parallel}^{(1)}\frac{\da
v_{\parallel}^{(1)}}{\da\theta}\right)\right],
\end{multline}
\begin{multline}\label{eq:secordgaslaw}
\frac{\gamma T_{0_c}}{c_{S_c}^2}p^{(2)}-T_{0_c}\rho^{(2)}
-\rho_{0_c}T^{(2)}=\\
\frac{\rho_{0_c}V}{c_{S_c}^2\cos\alpha}\left[\left(\frac{dT_0}{dx}\right)_c
+\frac{(\gamma-1)T_{0_c}}{\rho_{0_c}}\left(\frac{d\rho_0}{dx}\right)\right]\xi
v_{\parallel}^{(1)}\\
+\frac{(\gamma-1)T_{0_c}\rho_{0_c}v_{A_c}^2}{c_{S_c}^2\left(v_{A_c}^2+c_{S_c}^2\right)}
\left(v_{\parallel}^{(1)}\right)^2,
\end{multline}
\begin{multline}\label{eq:secordtotalpressure}
p^{(2)}+\frac{B_{0_c}}{\mu}B_{\parallel}^{(2)}=\frac{\rho_{0_c}V}{B_{0_c}\cos\alpha}
\left(\frac{dB_0}{dx}\right)_c\xi v_{\parallel}^{(1)}\\
-\frac{1}{3}\eta_0\cos\alpha\frac{3c_{S_c}^2+2v_{A_c}^2}{v_{A_c}^2+c_{S_c}^2}\frac{\da
v_{\parallel}^{(1)}}{\da\theta}-\frac{\rho_{0_c}c_{S_c}^2}{2\left(v_{A_c}^2+c_{S_c}^2\right)}\left(
v_{\parallel}^{(1)}\right)^2.
\end{multline}
In deriving the above system we have used
the fact that
\begin{equation}\label{eq:Qlin1}
Q^{(1)}=\frac{3c_{S_c}^2+2v_{A_c}^2}{v_{A_c}^2+c_{S_c}^2} \frac{\da
v_{\parallel}^{(1)}}{\da\theta}\cos\alpha,
\end{equation}
obtained from Eq. (\ref{eq:Q1}) in the first order of approximation. With the
exception of Eq. (\ref{eq:secordparainduct}), which has the
addition of the Hall term, these equations are identical to those
found by Ballai \textit{et al.} \cite{Ballai1998a}.

The left-hand sides of the set of Eqs.
(\ref{eq:secordconserve})-(\ref{eq:secordtotalpressure}) could be
obtained from the left-hand sides of the first order approximation
by substituting variables with the superscript `2' for those with
superscript `1'. The first order of approximation possesses a
non-trivial solution, so Eqs.
(\ref{eq:secordconserve})-(\ref{eq:secordtotalpressure}) are
compatible only if the right-hand sides of Eqs.
(\ref{eq:secordconserve})-(\ref{eq:secordtotalpressure}) satisfy a
compatibility condition. To derive the compatibility condition we
express $\rho^{(2)}$ and $B_{\parallel}^{(2)}$ in terms of
$u^{(2)}$, $v_{\parallel}^{(2)}$, $u^{(1)}$, $v_{\parallel}^{(1)}$
and $\widetilde{P}^{(1)}$, using Eqs. (\ref{eq:secordparamoment})
and (\ref{eq:secordthermal})-(\ref{eq:secordtotalpressure}).
Subsequently, we substitute these expressions into Eqs.
(\ref{eq:secordconserve}) and (\ref{eq:secordparainduct}), to obtain
\begin{multline}\label{eq:identicaleq1}
\frac{\da u^{(2)}}{\da\xi}+\frac{V^2}{v_{A_c}^2\cos\alpha}\frac{\da
v_{\parallel}^{(2)}}{\da\theta}=\\
\frac{V\left(v_{A_c}^2+c_{S_c}^2\sin^2\alpha\right)}{\rho_{0_c}v_{A_c}^4\cos^2\alpha}
\frac{d\widetilde{P}^{(1)}}{d\theta}
+\frac{V}{v_{A_c}^2+c_{S_c}^2}v_{\parallel}^{(1)}\frac{\da
v_{\parallel}^{(1)}}{\da\theta}\\
+\frac{V^2}{v_{A_c}^2\cos\alpha}\left[\frac{2}{B_{0_c}}\left(\frac{dB_0}{dx}\right)_c
-\frac{1}{\rho_{0_c}}\left(\frac{d\rho_0}{dx}\right)_c\right]\xi\frac{\da
v_{\parallel}^{(1)}}{\da\theta}-\\
\frac{\eta_0
V\cos\alpha}{\rho_{0_c}v_{A_c}^2}\left(\frac{2v_{A_c}^2+3c_{S_c}^2}{v_{A_c}^2+c_{S_c}^2}\right)
\frac{\da^2
v_{\parallel}^{(1)}}{\da\theta^2}-\frac{V}{c_{S_c}^2\cos\alpha}u^{(1)}\frac{\da
v_{\parallel}^{(1)}}{\da\xi}\\
-\frac{\chi\sin\alpha}{v_{A_c}^2+c_{S_c}^2}\frac{\da
v_{\parallel}^{(1)}}{\da\xi}\frac{\da
v_{\parallel}^{(1)}}{\da\theta},
\end{multline}
\begin{multline}\label{eq:identicaleq2}
\frac{\da u^{(2)}}{\da\xi}+\frac{V^2}{v_{A_c}^2\cos\alpha}\frac{\da
v_{\parallel}^{(2)}}{\da\theta}=\frac{c_{S_c}^2\sin^2\alpha}{\rho_{0_c}Vv_{A_c}^2}
\frac{d\widetilde{P}^{(1)}}{d\theta}\\
-\frac{V^2}{T_{0_c}c_{S_c}^2\cos\alpha}\left(\frac{dT_0}{dx}\right)_c\xi\frac{\da
v_{\parallel}^{(1)}}{\da\theta}-\frac{V}{c_{S_c}^2\cos\alpha}u^{(1)}\frac{\da
v_{\parallel}}{\da\xi}\\
-\frac{V\left(2c_{S_c}^2+(\gamma+1)v_{A_c}^2\right)}{c_{S_c}^2
\left(v_{A_c}^2+c_{S_c}^2\right)}v_{\parallel}^{(1)}\frac{\da
v_{\parallel}^{(1)}}{\da\theta}\\
+\frac{V\cos\alpha}{\gamma\rho_{0_c}c_{S_c}^2}
\left[\frac{2\gamma\eta_0}{3}\left(\frac{2v_{A_c}^2+3c_{S_c}^2}{v_{A_c}^2+c_{S_c}^2}\right)
+\frac{(\gamma-1)^2\kappa_{\parallel}}{\widetilde{R}}\right]\frac{\da^2
v_{\parallel}^{(1)}}{\da\theta^2}.
\end{multline}
It can be seen that Eqs. (\ref{eq:identicaleq1}) and
(\ref{eq:identicaleq2}) have identical left-hand sides. Extracting
these two equations we derive the compatibility condition, which is the
equation connecting $v_{\parallel}^{(1)}$ and $\widetilde{P}^{(1)}$
\begin{multline}\label{eq:governing1}
\Delta\xi\frac{\da
v_{\parallel}^{(1)}}{\da\theta}-av_{\parallel}^{(1)}\frac{\da
v_{\parallel}^{(1)}}{\da\theta}+\frac{V^3\lambda}{v_{A_c}^2+c_{S_c}^2}\frac{\da^2
v_{\parallel}^{(1)}}{\da\theta^2}\\
+\Omega\frac{\da v_{\parallel}^{(1)}}{\da\xi}\frac{\da
v_{\parallel}^{(1)}}{\da\theta}=\frac{Vc_{S_c}^2\cos\alpha}{\rho_{0_c}\left(v_{A_c}^2+c_{S_c}^2\right)}
\frac{d\widetilde{P}^{(1)}}{d\theta},
\end{multline}
where we have used the notation
\begin{align}
&a=\frac{V\left[(\gamma+1)v_{A_c}^2+3c_{S_c}^2\right]v_{A_c}^2\cos\alpha}
{\left(v_{A_c}^2+c_{S_c}^2\right)^2},\\
&\lambda=\frac{\eta_0\left(2v_{A_c}^2+3c_{S_c}^2\right)^2}
{3\rho_{0_c} v_{A_c}^2 c_{S_c}^2}+\frac{(\gamma-1)^2
\kappa_{\parallel}\left(v_{A_c}^2+c_{S_c}^2\right)}{\gamma\rho_{0_c}\widetilde{R}c_{S_c}^2},\\
&\Omega=\frac{{\chi}c_{S_c}^2v_{A_c}^2}{\left(v_{A_c}^2+c_{S_c}^2\right)^2}\cos\alpha\sin\alpha,\\
&\Delta=\left(\frac{dc_T^2}{dx}\right)_c.
\end{align}
Equation (\ref{eq:governing1}) differs from its counterpart found by
Ballai \textit{et al.} \cite{Ballai1998a} only by the last term of the left-hand side
representing Hall dispersion. If we linearize Eq.
(\ref{eq:governing1}) and take $v_{\parallel}^{(1)}$ proportional to
$\exp(ik\theta)$, we arrive at the linear equation for the parallel
velocity obtained in linear theory by Ruderman and Goossens
\cite{Ruderman1996}.

Equation (\ref{eq:governing1}) is the complete nonlinear governing
equation for the parallel velocity in the slow dissipative layer.
The function $\widetilde{P}^{(1)}$ in this equation is determined by
the solution outside the dissipative layer, and is thought to be the
driving term. We should note here that the dispersion (the last term
on the left-hand side) appears as a nonlinear term (\emph{nonlinear} dispersion).

\section{Nonlinear Connection Formulae}

In linear dissipative MHD it was assumed that when dissipative
effects are weak they are only important in the thin dissipative
layer that embraces the ideal resonant position (see, e.g.,
\cite{sakurai1}; \cite{hollweg1988b};
\cite{goossens1995};\cite{hollweg1988a}). Outside this layer, ideal
MHD can be employed to describe the plasma motion. The dissipative
layer is treated as a surface of discontinuity. In order to solve
Eq. (\ref{eq:linearequation}) boundary conditions are needed for the
variables $u$ and $P$ at this surface of discontinuity. In linear
theory these conditions are described by the explicit connection
formulae that determine the jumps in the quantities $u$ and $P$. In
order to derive the nonlinear counterpart of connection formulae we
first define the jump of a function, $f(x)$, across the dissipative
layer as
\begin{equation}\label{eq:jump}
\left[f\right]=\lim_{x \to +0}\left\{f(x)-f(-x)\right\},
\end{equation}
where the coordinate system has be translated such that $x_c=0$. The
thickness of this dissipative layer, $\delta_c$, is determined by
the condition that the first and third terms in Eq.
(\ref{eq:governing1}) are of the same order, i.e.
\begin{equation}\label{eq:dissipative1}
\delta_c=\frac{V^{3}k\lambda}{\left(v_{A_c}^2+c_{S_c}^2\right)|\Delta|},
\end{equation}
where $k$ is the wavenumber of waves. It is instructive to introduce a new,
dimensionless variable, $\sigma=x/\delta_c$, in the dissipative
layer. Let $x_0$ be the characteristic width of the overlap regions
of the dissipative layer (where both the linear ideal MHD equations and
the nonlinear dissipative MHD equations are valid).
One of the main reasons we have introduced the variable $\sigma$ is
the property that $\sigma=\mathscr{O}(1)$ in the dissipative layer,
while $|x|\rightarrow x_0$ corresponds to
$|\sigma|\rightarrow\infty$. This provides us with the second definition of the jump in
the function $f(x)$ across the dissipative layer,
\begin{equation}\label{eq:jump2}
\left[f\right]=\lim_{\sigma \to +\infty}\left\{f(\sigma)-f(-\sigma)\right\}.
\end{equation}
The first connection formula can be obtained in a straightforward
way by taking into account that the variable $\widetilde{P}^{(1)}$
does not change across the dissipative layer, so there cannot be any
jump in the total pressure,
\begin{equation}\label{eq:totalpressurejump}
\left[\bar{P}\right]=0.
\end{equation}
This connection formula is the same as obtained previously by linear
and nonlinear theories.

In order to derive the second connection formula we use the
approximate relations $u\approx \epsilon u^{(1)}$,
$v_{\parallel}\approx\epsilon^{1/2}v_{\parallel}^{(1)}$,
$\widetilde{P}\approx\epsilon\widetilde{P}^{(1)}$ and introduce the
new dimensionless variable, $q$, defined as
\begin{equation}\label{eq:q}
q=\epsilon^{1/2}\frac{kV\delta_c\cos\alpha}{v_{A_c}^2}v_{\parallel}^{(1)}.
\end{equation}
In the new variable, Eqs. (\ref{eq:relations5}) and
(\ref{eq:governing1}) are rewritten as
\begin{equation}\label{eq:uqrelation1}
\frac{\da u}{\da\sigma}=-\frac{V}{k\cos^2\alpha}\frac{\da
q}{\da\theta},
\end{equation}
\begin{multline}\label{eq:governing2}
\mbox{sign}(\Delta)\sigma\frac{\da q}{\da\theta}-\Lambda q\frac{\da q}{\da\theta}
+k^{-1}\frac{\da^2 q}{\da\theta^2}\\
+\Psi\frac{\da q}{\da\sigma}\frac{\da q}{\da\theta}=
\frac{kV^4}{\rho_{0_c}v_{A_c}^2|\Delta|}\frac{d\widetilde{P}}{d\theta},
\end{multline}
where
\begin{align}
&\Lambda=R^2\frac{v_{A_c}^4|\Delta|\left[(\gamma+1)v_{A_c}^2+3c_{S_c}^2\right]}{kV^8},\label{eq:lambda}\\
&\Psi=R^2\frac{\chi|\Delta|^2 c_{S_c}^2 v_{A_c}^2
(v_{A_c}^2+c_{S_c}^2)\sin\alpha}{kV^{13}}.\label{eq:psi}
\end{align}
We have used a slightly different Reynolds number, $R$, to the one
used in previous sections, here it is defined as
\begin{equation}\label{eq:newreynold}
R=\frac{V}{k\lambda}.
\end{equation}
It is easy to show that the estimations
\begin{align*}
&q=\mathscr{O}(\epsilon^{1/2}kl_{inh}R^{-1}),\\
&\delta_c=\mathscr{O}(l_{inh}R^{-1}),\\
&\Lambda=\mathscr{O}(R^2 k^{-1}l_{inh}^{-1}),\\
&\Psi=\mathscr{O}(R^2 k^{-1}l_{inh}^{-2}),
\end{align*}
are valid. The ratios of the nonlinear to dissipative term, dispersive to dissipative term, and dispersive to nonlinear term in Eq.
(\ref{eq:governing2}), $N$, $D_{d}$ and $D_{n}$ respectively, are
\begin{equation}\label{eq:N}
N=\mathscr{O}(\epsilon^{1/2}Rkl_{inh}),
\end{equation}
\begin{equation}\label{eq:Dd}
D_{d}=\mathscr{O}(\epsilon^{1/2} R^2 kl_{inh}^{-1}),
\end{equation}
\begin{equation}\label{eq:Dn}
D_{n}=\mathscr{O}(Rl_{inh}^{-2}).
\end{equation}
The parameters $N$ and $D_d$ can be considered as nonlinearity and
dispersive parameters, respectively. Nonlinearity (dispersion) is
important if $N\gtrsim 1$ ($D_d\gtrsim 1$). When $N\ll 1$
($D_d\ll1$) the nonlinear (dispersive) term in Eq.
(\ref{eq:governing2}) can be neglected. Dispersion dominates
nonlinearity if $D_n>1$. In the opposite case, $D_n<1$, nonlinearity
dominates dispersion. When $kl_{inh}=\mathscr{O}(1)$ (in fact our
analysis is valid when $\epsilon^{1/2}\ll kl_{inh}\ll
\epsilon^{-1/2}$), the total Reynolds number, $R$, used in this
section is of the same order of magnitude as that used in the
previous sections, and the criterion of nonlinearity coincides with
that obtained in section II from the qualitative analysis. With the
above scalings in mind, it is obvious that the physical background
of this paper is applicable for relatively short inhomogeneity
scales.

Following linear studies, the outer solution reveals that
$v_{\parallel}=\mathscr{O}(x^{-1})$ as $x\rightarrow 0$. Thus, to
match the outer and inner solutions in the overlap regions the
asymptotic relation $q=\mathscr{O}(\sigma^{-1})$ as
$|\sigma|\rightarrow \infty$ must be valid. It then directly follows
from Eq. (\ref{eq:governing2}) that
\begin{equation}\label{eq:q1}
q\simeq\frac{kV^4\widetilde{P}_c(\theta)}{\rho_{0_c}v_{A_c}^2\Delta\sigma},
\end{equation}
for $|\sigma|\rightarrow \infty$. From Eqs. (\ref{eq:uqrelation1})
and (\ref{eq:q1}), we obtain that
\begin{equation}\label{eq:ufinal}
u=-\frac{Vc_{S_c}^4\cos^2\alpha}{\rho_{0_c}\Delta(v_{A_c}^2+v_{A_c}^2)^2}
\frac{d\widetilde{P}_c}{d\theta}\ln|\sigma|+u_{\pm}(\theta)+\mathscr{O}(\sigma^{-1})
\end{equation}
for $\sigma\rightarrow\pm\infty$. The functions $u_{+}$ and $u_{-}$ are related by
\begin{equation}\label{eq:upm}
u_{+}(\theta)-u_{-}(\theta)=
-\frac{V}{k\cos^2\alpha}\mathscr{P}\int_{-\infty}^{\infty}\frac{\da
q}{\da\theta}\phantom{.}d\sigma.
\end{equation}
Equation (\ref{eq:upm}) uses the symbol of Cauchy principal part,
$\mathscr{P}$, because the integral is divergent at
infinity. So, in accordance with Eq. (\ref{eq:jump2}), we obtain the
implicit jump condition for the normal component of velocity
\begin{equation}\label{eq:jumpforu}
\left[u\right]=-\frac{V}{k\cos^2\alpha}\mathscr{P}\int_{-\infty}^{\infty}\frac{\da
q}{\da\theta}\phantom{.}d\sigma.
\end{equation}

Equation (\ref{eq:jumpforu}) is the nonlinear analog for the
implicit connection formula for the normal component of velocity.
The main difference between the linear and the nonlinear connection
formula is that while in the linear version the jumps are expressed
explicitly in terms of $\widetilde{P}(\theta)$ and equilibrium
quantities, in the nonlinear version the jump in the normal
component of velocity, $u$, is expressed implicitly in terms of an
unknown quantity $q$. The connection formula (\ref{eq:jumpforu}) is
identical to that found by Ballai \textit{et al.} \cite{Ballai1998a}
and Ruderman \textit{et al.} \cite{ruderman3} in the limit of weak
nonlinearity for non-dispersive plasmas. To find solutions in the
dissipative layer we have to use Eqs. (\ref{eq:linearequation}) and
(\ref{eq:governing2}) simultaneously. The boundary conditions for
the outer solution are provided by Eqs. (\ref{eq:totalpressurejump})
and (\ref{eq:jumpforu}).

The governing equation for $q$, Eq. (\ref{eq:governing2}), differs to
that derived by Ballai \textit{et al.} \cite{Ballai1998a} only in
the appearance of the last term in the left-hand side which is
related to the consideration of the Hall term in the generalized
Ohm's law.

\section{Conclusions}

In the present paper we have further developed the nonlinear theory
of resonant slow MHD waves in the dissipative layer in
one-dimensional planar geometry in plasmas with strongly anisotropic
viscosity and thermal conductivity by considering dispersive effects. The plasma motion outside
the dissipative layer is described by the set of linear, ideal MHD
equations. This set of equations can be reduced to Eq.
(\ref{eq:linearequation}) for the component of the velocity in the
direction of the inhomogeneity, $u$, and the perturbation of total
pressure, $P$. The wave motion in the dissipative layer is governed
by Eq. (\ref{eq:governing2}) for the quantity $q$, which is the
dimensionless component of the velocity parallel to the equilibrium
magnetic field, defined by Eq. (\ref{eq:q}). The dissipative layer
is considered as a surface of discontinuity when solving Eq.
(\ref{eq:linearequation}) to describe the wave motion outside the
dissipative layer. The jumps across the dissipative layer are given
by Eqs. (\ref{eq:totalpressurejump}) and (\ref{eq:jumpforu}), thus
providing the boundary conditions at the surface of discontinuity.
In stark contrast to linear theory, the jump in $u$ is not solvable
analytically, as it given in terms of a infinite integral of $q$ -
which in turn is determined by Eq. (\ref{eq:governing2}). Since this
equation has not been solved analytically we must solve Eqs.
(\ref{eq:linearequation}) and (\ref{eq:governing2}) simultaneously
when studying resonant slow waves that are nonlinear in the
dissipative layer.

The reader should note that in the upper solar corona the plasma
$\beta$ (ratio of kinetic to magnetic pressure) is very small, hence
the significance of slow resonance is dramatically reduce (in the
limit $\beta=0$, slow waves cease to exist). The best applicability
for the present paper is in the regions of the chromosphere and
lower solar corona.

It is interesting to note that the dispersion at the slow resonance
appears in the form of a nonlinear term (\emph{nonlinear}
dispersion). It is expected that if the Hall term is included in the
Alfv\'{e}n resonance, this will appear as a linear term. The
governing equations and the jump conditions will be used later in
studying the absorption of an external driver in the limit of weak
and strong nonlinearity. It is intended that the authors will
investigate the possibility of describing the propagation of
solitary and shock waves in the slow dissipative layer. It remains
to be investigated how the inclusion of the dispersive term will
influence the conclusions drawn when resonant absorption was used to
explain the damping of waves and oscillations in the coronal
seismology framework.

Earlier studies (see, e.g. \cite{Ruderman1997}; \cite{Ballai1998b}),
show that nonlinearity decreases the effective absorption of waves
in the limit of weak nonlinearity. The effect of dispersion on the
absorption of waves in slow dissipative layers will be addressed in
the near future.

\section*{ACKNOWLEDGEMENTS}

The authors would like to thank STFC (Science and Technology
Facilities Council) for the financial support provided. Authors
would like to acknowledge Prof. M. S. Ruderman for his tireless help
and advice with the work carried out in this paper. I. Ballai
acknowledges the financial support by NFS Hungary (OTKA, K67746) and
The National University Research Council Romania
(CNCSIS-PN-II/531/2007).

\section*{APPENDIX: THE DERIVATION OF THE HALL TERM IN THE INDUCTION EQUATION FOR SLOW RESONANT WAVES IN THE DISSIPATIVE LAYER}

In this appendix we will derive the components of the Hall term in
the induction equations and study the conditions under which this extra
effect is important. The parallel component of the magnetic field
perturbation dominates the other components in the slow dissipative
layer. The Hall term contains the first derivative of this parallel
component of the magnetic field perturbation with respect to $z$.
The first term of Braginskii's viscosity tensor contains the second
derivative of the parallel component of the magnetic field
perturbation with respect to $z$. As a result the Hall term can be of
the same order or larger than the dissipative term.

The generalized Ohm's law including the Hall term can be written as
(see, e.g., \cite{priest1})
\begin{equation}\tag{A1}\label{eq:ohmslaw}
{\bf E}=-{\bf v}\times{\bf B}+\frac{1}{\sigma}{\bf
j}+\frac{1}{en_e}{\bf j}\times{\bf B},
\end{equation}
where ${\bf E}$ is the electric field, ${\bf j}$ the density of the
electrical current, $n_e$ the electron number density, $e$ the
electron charge and $\sigma$ the electrical conductivity. The
density of electrical current and magnetic induction, ${\bf B}$, are
related by Amp\`{e}re's law
\begin{equation}\tag{A2}\label{eq:ampereslaw}
{\bf j}=\frac{1}{\mu}\nabla\times{\bf B},
\end{equation}
with the electrical conductivity given by
\begin{equation}\tag{A3}\label{eq:eleccon1}
\sigma=\frac{n_{e}e^2 m_{e}^{-1}}{\tau_{e}^{-1}+\tau_{n}^{-1}}.
\end{equation}
Here $m_e$ is the electron mass, $\tau_{e}$ the electron
collision time and $\tau_{n}$ the neutral collision time.

For a fully-ionized, collision-dominated, plasma Eq.
(\ref{eq:eleccon1}) reduces to
\begin{equation}\tag{A4}\label{eq:eleccon2}
\sigma\approx\frac{n_{e}e^2\tau_{e}}{m_{e}}.
\end{equation}
In accordance with Spitzer \cite{spitzer1962} the electron
collision time is given by
\begin{equation}\tag{A5}\label{eq:eitime}
\tau_{e}=2.66\times 10^{5}\frac{T^{3/2}}{n_{e}\ln\Lambda}\mbox{s},
\end{equation}
where $T$ is the temperature and $\ln\Lambda$ is the Coulomb
logarithm (here taken to be $22$). From Eq. (\ref{eq:eitime}),
$\tau_{e}$ changes from $9.4\times10^{-8}\mbox{s}$ in the upper
photosphere to $1.4\times 10^{-2}\mbox{s}$ in the solar corona. On
the other hand, $\omega_e$ changes from $1.8\times
10^{10}\mbox{s}^{-1}$ in the upper photosphere to $1.8\times
10^{8}\mbox{s}^{-1}$ in the solar corona. As a consequence the Hall
parameter, $\omega_e\tau_{e}$, changes from $1.69\times10^3$ in the
upper photosphere to $2.52\times10^{6}$ in the solar corona. Since
$\omega_e\tau_{e}\gg1$, the Hall term cannot be neglected in the
upper photosphere nor the solar corona.

In order to estimate the relative importance of the Hall term and
viscous term in the dissipative layer we must employ a more
sophisticated analysis similar to the analysis presented by Ruderman \textit{et al.} \cite{ruderman3}.
The generalized induction equation (neglecting finite electrical
resistivity) is,
\begin{equation}\tag{A6}\label{eq:genindeq}
\frac{\da\overline{\mathbf{B}}}{\da
t}=\nabla\times(\mathbf{v}\times\overline{\mathbf{B}})+\frac{1}{\mu
e}\nabla\times
\left(\frac{1}{n_e}\overline{\mathbf{B}}\times\nabla\times\overline{\mathbf{B}}\right).
\end{equation}
In what follows we assume that the ionization coefficient is
constant, so that $n_e$ is proportional to $\bar{\rho}$, and in
particular $n_{e}^{-1}\nabla n_e=\bar{\rho}^{-1}\nabla\bar{\rho}$.
Equations (\ref{eq:linearexpansion}) and (\ref{eq:innerexpansion})
provide the following estimations in the dissipative layer:
\begin{align}\label{eq:estimates}
&u=\mathscr{O}(\epsilon),\phantom{x}v_{\perp}=\mathscr{O}(\epsilon),\phantom{x}
\rho=\mathscr{O}(\epsilon^{1/2}),\nonumber\\
&v_{\parallel}=\mathscr{O}(\epsilon^{1/2}),\phantom{x}
B_{\parallel}=\mathscr{O}(\epsilon^{1/2}),\tag{A7}
\end{align}
where $\epsilon$ still denotes the dimensionless amplitude of
oscillations far away from the dissipative layer.

The thickness of the dissipative layer divided by the characteristic
scale of inhomogeneity is
$\delta_c/l_{inh}=\mathscr{O}(\epsilon^{1/2})$. This gives rise to
\begin{align}\label{eq:inhomoq}
&l_{inh}\frac{\da h}{\da
x}=\mathscr{O}(\epsilon^{-1/2}h),\phantom{x}
l_{inh}\frac{\da h}{\da z}=\mathscr{O}(h),\nonumber\\
&l_{inh}^{2}\frac{\da^2 h}{\da z^2}=\mathscr{O}(h),\tag{A8}
\end{align}
where $h$ denotes any of the quantities $u$, $\rho$,
$v_{\parallel}$ or $B_{\parallel}$. Since the first term in the
expansion of $B_\perp$ is independent of $x$, it follows that
\begin{align}\label{eq:inhomobperp}
&l_{inh}\frac{\da B_{\perp}}{\da
x}=\mathscr{O}(B_{\perp}),\phantom{x}
l_{inh}\frac{\da B_{\perp}}{\da z}=\mathscr{O}(B_{\perp}),\nonumber\\
&l_{inh}^{2}\frac{\da^2 B_{\perp}}{\da
x^2}=\mathscr{O}(\epsilon^{-1/2}B_{\perp}),\tag{A9}
\end{align}
(the same applies to the variable $v_{\perp}$).

We now need to calculate the components of the vectors of the
Braginskii's viscosity tensor and the Hall term from Eq.
(\ref{eq:genindeq}) normal to the magnetic surfaces (the
$x$-direction) and in the magnetic surfaces parallel and
perpendicular to the equilibrium magnetic field lines. We use Eqs.
(\ref{eq:inhomoq}) and (\ref{eq:inhomobperp}) in order to estimate
all the terms and we only retain the largest. The components of the
Braginskii tensor acting in the normal and perpendicular directions
relative to the equilibrium magnetic field are the second and third
ones, describing shear viscosity (even though in the paper we only
consider Eq. (\ref{eq:braginskii}) we need the second and third
terms of Braginskii's viscosity tensor to complete our scalings).
Since they are of the same order, for the purpose of our estimations
it is enough to consider $\overline{\eta}_1$ only. Braginskii's
viscosity tensor simplifies to,
\begin{align}
&\overline{\eta}_1(\nabla\cdot
S_1)_x=\overline{\eta}_1\frac{\da^2 u}{\da x^2}+\ldots,\tag{A10}\\
&\overline{\eta}_1(\nabla\cdot
S_1)_{\perp}=\overline{\eta}_1\frac{\da^2 v_{\perp}}{\da x^2}+\ldots,\tag{A11}\\
&\overline{\eta}_0(\nabla\cdot
S_0)_{\parallel}=\overline{\eta}_0\cos\alpha\left(2\cos\alpha\frac{\da^2
v_{\parallel}}{\da z^2}-\frac{\da^2 u}{\da x\da
z}\right)+\ldots.\tag{A12}
\end{align}
It should be stated that $\overline{\eta}_0\gg\overline{\eta}_1$ and
$\overline{\eta}_0(\nabla\cdot S_1)_x=\overline{\eta}_0(\nabla\cdot
S_1)_{\perp}=0$. The components of the Hall term in the induction
equation reduce to
\begin{align}
&H_x=\frac{B_0\cos\alpha\sin\alpha}{\mu en_e}\frac{\da^2 B_{\parallel}}{\da z^2}+\ldots,\tag{A13}\\
&H_{\perp}=\frac{B_0\cos\alpha}{\mu en_e}\frac{\da^2 B_{\parallel}}{\da x\da z}+\ldots,\tag{A14}\\
&H_{\parallel}=\frac{B_0\sin{\alpha}}{\rho_0\mu en_e}\frac{\da
B_{\parallel}}{\da z}\frac{\da\rho}{\da x}+\ldots.\tag{A15}
\end{align}

With the aid of Eqs. (\ref{eq:estimates}), (\ref{eq:inhomoq}) and
(\ref{eq:inhomobperp}) we obtain the ratios
\begin{align}
&\frac{H_x}{\overline{\eta}_1(\nabla\cdot
S_1)_x}\sim\epsilon^{1/2}\frac{\overline{\chi}}{\overline{\eta}_1},\tag{A16}\label{eq:xrat}\\
&\frac{H_{\perp}}{\overline{\eta}_1(\nabla\cdot
S_1)_{\perp}}\sim\epsilon^{-1/2}\frac{\overline{\chi}}{\overline{\eta}_1},\tag{A17}\label{eq:perprat}\\
&\frac{H_{\parallel}}{\overline{\eta}_0(\nabla\cdot
S_0)_{\parallel}}\sim\frac{\overline{\chi}}{\rho_0\overline{\eta}_0}.\tag{A19}\label{eq:pararat}
\end{align}
Where $\overline{\chi}=\overline{\eta}\omega_e\tau_e$ is the coefficient of Hall conduction and $\overline{\eta}=1/{\sigma\mu}$ is the magnetic diffusivity. Strictly speaking, even the diffusivity is anisotropic in the solar
corona, but the parallel and perpendicular components only differ by
a factor of $2$. It has been noted that magnetic diffusivity is much
much smaller that the compressional viscosity in the solar corona.
However, in the coefficient of Hall conduction, $\overline{\chi}=\overline{\eta}\omega_e\tau_e$, we observe that the magnetic diffusivity
is multiplied by the product $\omega_e\tau_e$, which is very large in
the solar corona ($10^4-10^6$). Moreover, if we look at Eq.
($\ref{eq:pararat}$) we see that the coefficient of Hall conduction is divided by the density, which
is very small under solar coronal conditions. Therefore, the
parallel component of the Hall term in the induction equation
becomes very important in the slow dissipative layer.

The Hall terms in the normal and perpendicular direction relative to
the background magnetic field are included for completeness in the
paper, but they do not play a role in the governing equation (i.e.
can be left out completely and will not alter the result shown).
This is attributed to the fact that the dominant dynamics of
resonant slow waves is in the parallel direction relative to the
ambient magnetic field.

In summary, the Hall term in the parallel direction relative to the
ambient magnetic field, $H_{\parallel}$, must be included in the
slow dissipative layer when $\omega_e\tau_{e}\gg1$ because it is the
same order of magnitude (or larger) than the compressional viscous
term.

\newpage

\end{document}